\documentclass[utf8]{FrontiersinHarvardNoLogo}

\usepackage{url,hyperref,lineno,microtype,subcaption}
\usepackage{amssymb}
\usepackage{amsmath}
\usepackage{hyperref}
\usepackage{upgreek}
\usepackage{svg}
\usepackage[onehalfspacing]{setspace}

    \def\firstAuthorLast{K.~Altenm\"{u}ller {et~al.}} 
    \def\Authors{K.~Altenm\"{u}ller\,$^{1,*}$, J.~F.~Castel\,$^{1}$, S.~Cebri\'an\,$^{1}$, T.~Dafni\,$^{1}$, D.~D\'iez-Iba\~{n}ez\,$^{1}$, A.~Ezquerro\,$^{1}$, E.~Ferrer-Ribas\,$^{2}$, J.~Galan\,$^{1}$, J.~Galindo\,$^{1,3}$, J.~A.~Garc\'ia\,$^{1}$, A.~Giganon\,$^{2}$, C.~Goblin\,$^{2}$, I.~G.~Irastorza\,$^{1}$, C.~Loiseau\,$^{2}$, G.~Luz\'on\,$^{1}$, 
    X.~F.~Navick\,$^{2}$, C.~Margalejo\,$^{1}$, H.~Mirallas\,$^{1}$, L.~Obis\,$^{1}$, A.~Ortiz de Sol\'{o}rzano\,$^{1}$, T.~Papaevangelou\,$^{2}$,  O.~P\'erez\,$^{1}$, A.~Quintana\,$^{1,2} $, J.~Ruz\,$^{1}$ and J.~K.~Vogel\,$^{1,4}$}
    \def\Address{$^{1}$Center for Astroparticles and High Energy Physics (CAPA), Universidad de Zaragoza, Zaragoza, Spain\\
    $^{2}$Université Paris-Saclay, CEA, IRFU, Gif-sur-Yvette, France\\
    $^{3}$Instituto Tecnológico de Aragón (ITA), Zaragoza, Spain\\
    $^{4}$LLNL Lawrence Livermore National Laboratory, Livermore, CA,USA
    }
    \def\corrAuthor{Centro de Astropartículas y Física de Altas Energías (CAPA), Facultad de Ciencias, Universidad de Zaragoza, Pedro Cerbuna 12, 50009 Zaragoza (Spain)}
    
    \def\corrEmail{konrad.altenmueller@unizar.es}

\begin{document}

\onecolumn
\firstpage{1}
	
\title{Background discrimination with a Micromegas detector prototype and veto system for BabyIAXO}

\author[\firstAuthorLast ]{\Authors} 
\address{} 
\correspondance{} 

\extraAuth{}

\maketitle

\begin{abstract}
	In this paper we present measurements performed with a Micromegas X-ray detector setup. The detector is a prototype in the context of the \mbox{BabyIAXO} helioscope, which is under construction to search for an emission of the hypothetical axion particle from the sun. An important component of such a helioscope is a low background X-ray detector with a high efficiency in the 1--10\,keV energy range.
	The goal of the measurement was to study techniques for background discrimination. In addition to common techniques we used a multi-layer veto system designed to tag cosmogenic neutron background.
	Over an effective time of 52 days, a background level of $8.6 \times 10^{-7} \, \textnormal{counts keV}^{-1} \, \textnormal{cm}^{-2} \, \textnormal{s}^{-1}$ was reached in a laboratory at above ground level. This is the lowest background level achieved at surface level. In this paper we present the experimental setup, show simulations of the neutron-induced background, and demonstrate the process to identify background signals in the data. Finally, prospects to reach lower background levels down to $10^{-7} \, \textnormal{counts keV}^{-1} \, \textnormal{cm}^{-2} \, \textnormal{s}^{-1}$  will be discussed.
\end{abstract}

	\section{Introduction}
	\label{sec:intro}
	\subsection{Axions as dark matter candidates}
	The measurements presented in this paper were performed with a prototype detector for the BabyIAXO helioscope. BabyIAXO will search for an emission of the hypothetical axion particle from the sun ~\citep{BabyIAXO,IAXO2}.
	The axion is a theoretically well motivated candidate particle for the dark matter. It arises as a solution to the strong CP problem through the Peccei-Quinn mechanism~\citep{PQ1,PQ2}. Generic axions couple to photons, which permits the transformation of axions and photons into each other via the Primakoff effect. Non-hadronic axion models predict additionally a significant coupling of axions to electrons~\citep[e.g.][]{Red1,Hoof:2021mld}. 
	After simple models that predict relatively heavy axions were ruled out by previous experiments, the axion -- if it exists -- has to be very light with a low interaction strength, which makes it a well suited dark matter candidate \citep{axions,DM1}. 
    Several anomalous astrophysical observations hint at the existence of axions: axion-photon oscillations could explain both the unexpectedly fast cooling of white dwarfs and the transparency of the universe to very high-energy gamma rays~\citep{Ast1,DiLuzio:2021ysg,Ast2}. The region of the parameter space motivated by both these hints is partially in reach for a next-generation helioscope like BabyIAXO~\citep{BabyIAXO}.
    BabyIAXO is a fully-fledged helioscope that will explore relevant parameter space, however, it also serves as a prototype for an even larger experiment -- IAXO, the \textbf{I}nternational \textbf{AX}ion \textbf{O}bservatory~\citep{IAXO1,IAXO2}. 
	At the same time, BabyIAXO is the successor of the CERN Axion Solar Telescope (CAST), which took data until the end of $2021$ and delivered current benchmark limits over a wide range in the axion parameter space, which are at the same level as astrophysical limits~\citep{CASTresult}.
	
	\subsection{The BabyIAXO axion helioscope}
	\label{sec:BabyIAXO}
	The principle of an axion helioscope is to use the inverse Primakoff effect to convert solar axions that reach earth into detectable photons~\cite{helioscope}. For this purpose, a strong magnet is aligned with the sun. Axions passing through the volume of the magnetic field, which contains a high density of virtual photons, can transform into X-ray photons. These photons are then focused by X-ray optics onto low-background detectors~\citep{IAXO1}.
	The device-specific characteristics that determine the discovery potential are the strength, length and volume of the magnet, the throughput of the optics and smallness of the focal spot, the efficiency and background level in the energy range of interest in the detector as well as the exposure time tracking the Sun~\citep{BabyIAXO}. In this paper, we are discussing only the final component in this chain, the detector.
 
	Computing the expected energy spectrum of solar axions (or the resulting photons) shows us the requirements of this detector. The spectrum of axions produced by the Primakoff effect is a distribution with a peak at $\sim 3$\,keV and a tail that extends to $\sim 10$\,keV. In other non-hadronic models, the contribution from the interaction of axions with electrons is dominating the spectrum and results in a peak at $\sim 1$\,keV~\citep{Red1}. Thus a detector with a low threshold and a high efficiency in the 1--10\,keV range is required. In addition, due to the rareness of axion-photon conversions, the background level has to be as low as possible. In the baseline design it is intended to use a Micromegas detector, which has been shown to achieve very low background levels~\citep{CASTresult}. Further detectors with different technological approaches are under development for later stages of the experiment~\citep{detectorsNIMA}.
 
\begin{figure}
		\centering
		\includegraphics[width=0.75\linewidth]{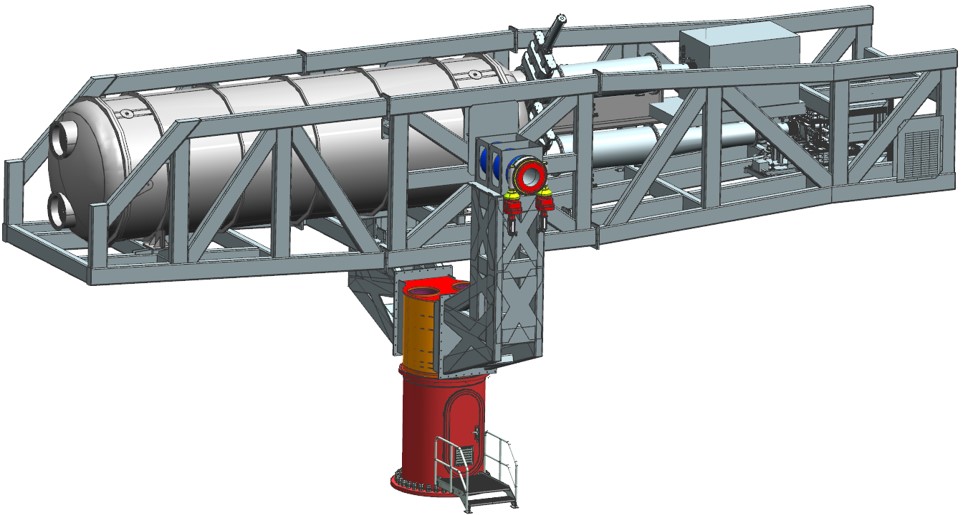}
		\caption{A sketch of the BabyIAXO helioscope. Mounted in a frame on the drive system, one can see the cryostat containing the magnet on the left, connected on the right to the X-ray telescopes and the detectors within their veto-systems. The total length of the helioscope is $\sim 21$\,m~\citep{BabyIAXO}.}
		\label{fig:BabyIAXO}
	\end{figure}

	A conceptual sketch of the current BabyIAXO design is shown in figure \ref{fig:BabyIAXO}. For more information, the design is presented in great detail in the paper~\citep{BabyIAXO}, while a full review of the physics case relevant for BabyIAXO can be found in the reference~\citep{IAXOphysics}. BabyIAXO is expected to be hosted by DESY in Hamburg, its construction has recently started.
 
    The work presented in this paper is about measurements with a Micromegas detector prototype for BabyIAXO, where a novel veto system was utilized to push the background down to an unprecedented level. In the next sections we describe the setup, show simulations of the setup, and then present methods to identify background events. Finally, we present and discuss the achieved background level from measurements in a laboratory at surface.
    
	\section{The IAXO-D0 detector prototype setup}
    \label{sec:setup}
\begin{figure}
		\centering
		\includegraphics[width=0.49\textwidth]{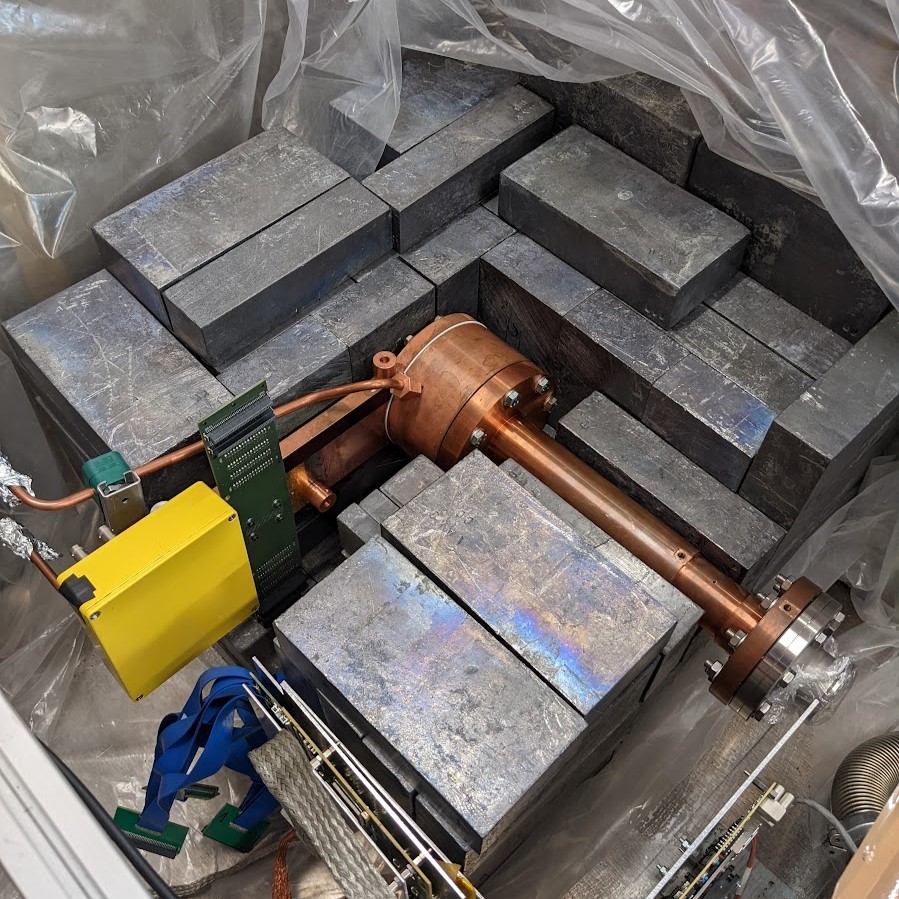}
		\hfill\includegraphics[width=0.49\textwidth]{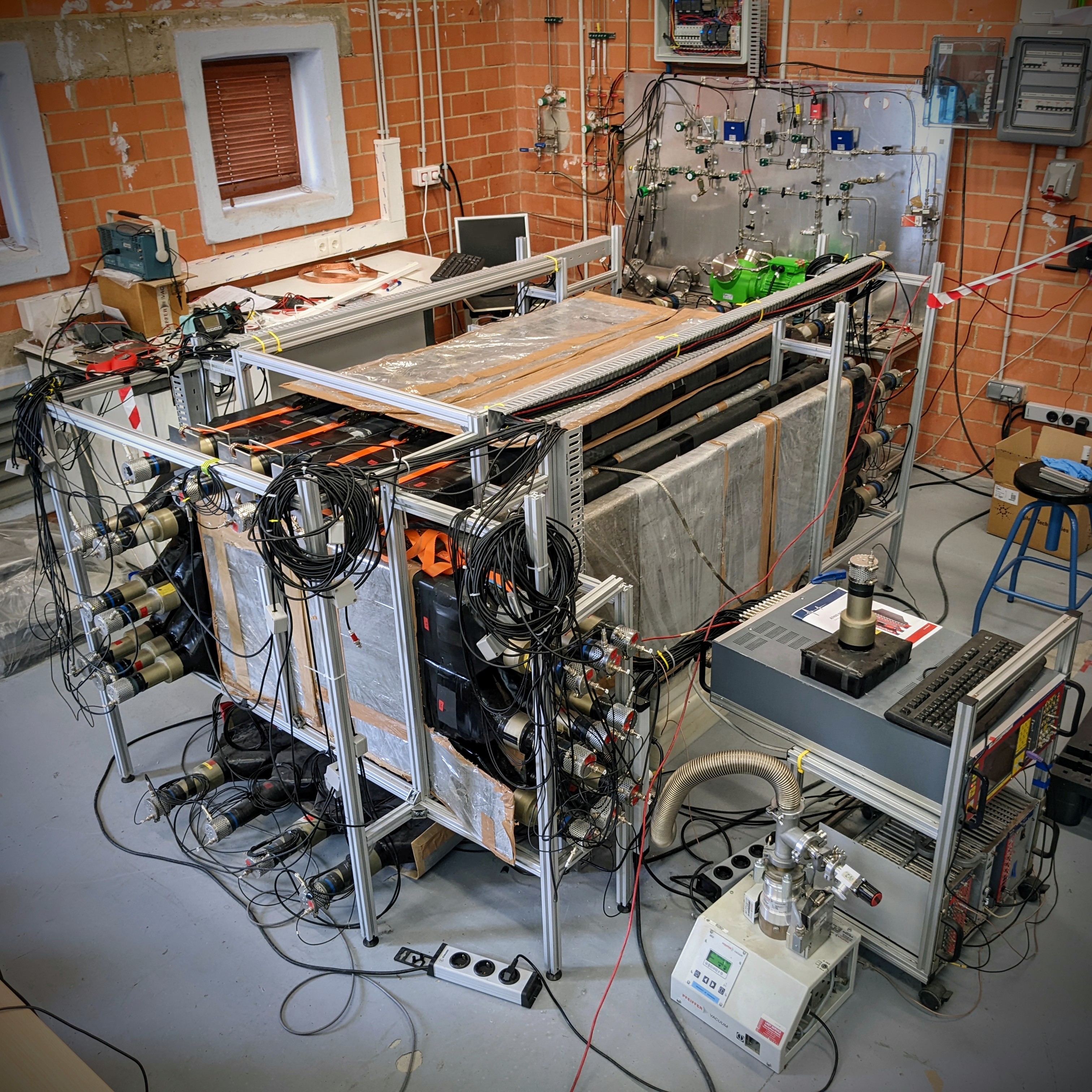}
		\caption{The IAXO-D0 setup. On the left the Micromegas detector within its lead castle. On the right the entire setup including the triple layer veto system.}
		\label{fig:setup}
	\end{figure}

	At CAPA\footnote{Centro de Astropart\'{i}culas y F\'{i}sica de Altas Energ\'{i}as, \url{https://capa.unizar.es/en/}} of the University of Zaragoza, a Micromegas prototype detector, dubbed IAXO-D0, was installed to take background data in order to characterize it under different conditions and to improve the background rejection. See figure \ref{fig:setup} for photos of the setup.
	Previously, with similar Micromegas detectors, background levels of $1 \times 10^{-6} \, \textnormal{counts keV}^{-1} \, \textnormal{cm}^{-2} \, \textnormal{s}^{-1}$ were demonstrated at surface level at CAST~\citep{CASTresult}.
	However, measurements in the underground laboratory at Canfranc with a CAST replica detector suggested much lower levels of $\sim 2 \times 10^{-7} \, \textnormal{counts keV}^{-1} \, \textnormal{cm}^{-2} \, \textnormal{s}^{-1}$ -- one order of magnitude lower than what was achieved at surface level despite using a muon veto \citep{MM_Canfranc}. It is suspected that the additional contribution at surface is caused by cosmic neutrons, a hypothesis supported by simulations, as explained in section~\ref{sec:sims}. 
	To shield the detector from natural background, the Micromegas is placed inside a lead shielding of 20\,cm thickness and surrounded by a 4-$\pi$ muon veto.
	In an attempt to discriminate additionally the background contribution by neutrons, the veto system consists of 3-layers of plastic scintillator panels with cadmium sheets in between the layers. Montecarlo simulations have shown that a primary neutron can interact in the setup, producing secondary neutrons that eventually get thermalized in the plastic scintillator panels of the vetoes. If thermal neutrons get captured in the cadmium sheets, gamma rays are produced that can be detected by the scintillators. In general, we expect that a neutron event produces many signals in the veto system over a time scale of several tens of $\upmu$s. This process is described in more detail in section \ref{sec:sims}.
 
	\subsection{The Micromegas detector}	
	The Micromegas detector is a gas-filled time projection chamber with a Micromesh Gaseous Structure (Micromegas) readout plane~\citep{Micromegas}. 
	The volume filled with gas has a length of 3\,cm. On one side, the volume is closed by an entrance window made of $4\,\upmu$m-thick aluminized mylar, that also serves as cathode at a voltage of $\sim -750$\,V. On the other end, the readout covers a surface of $6 \times 6$\,cm$^2$ with 120 strips in each direction. 
	The Micromegas readout of IAXO-D0 is produced via the \textit{microbulk} method using only radiopure copper and Kapton~\citep{microbulk}. It includes a mesh that is placed 50\,\textmu m above the strips and kept at a voltage between $-370$ and $-390$\,V. The detector is thus divided in two regions: a drift region between mesh and cathode, and an amplification region between the mesh and the strip readout, where the latter serves as anode at ground level. The volume is surrounded by rings at intermediate voltages that shape the electric field to improve the drift characteristics.

    In the context of the axion search, the energy range of interest (ROI) is 2--7\,keV. The lower limit of the energy range is set slightly above the detector threshold. The upper limit is set to 7\,keV, because above this energy the expected solar axion flux is negligible and one can evade signals from the 8\,keV copper K\textsubscript{$\upalpha$} X-ray fluorescence peak (visible in figure \ref{fig::MMspectrum}).
	Depending on the gas mixture and pressure used, a detection efficiency of 60--70\% is reached in the ROI. The detector threshold is $\sim 1$\,keV and mostly limited by the thickness of the entrance window.
	The gas used in the measurements presented here is a mixture of Xenon (48.85\%) + Neon (48.85\%) + C\textsubscript{4}H\textsubscript{10}  (2.3\%) at a pressure of 1050\,mbar. The isobutane (C\textsubscript{4}H\textsubscript{10}) serves as a quencher. A gas system recirculates the detector gas and allows to select pressure and flow.
	The system is calibrated with an \textsuperscript{55}Fe-source. The source is moved by a remotely controlled manipulator, connected to the evacuated volume in front of the entrance window.
	
	\subsection{The veto system}
	
\begin{figure}
		\centering
		\includegraphics[width=0.8\textwidth]{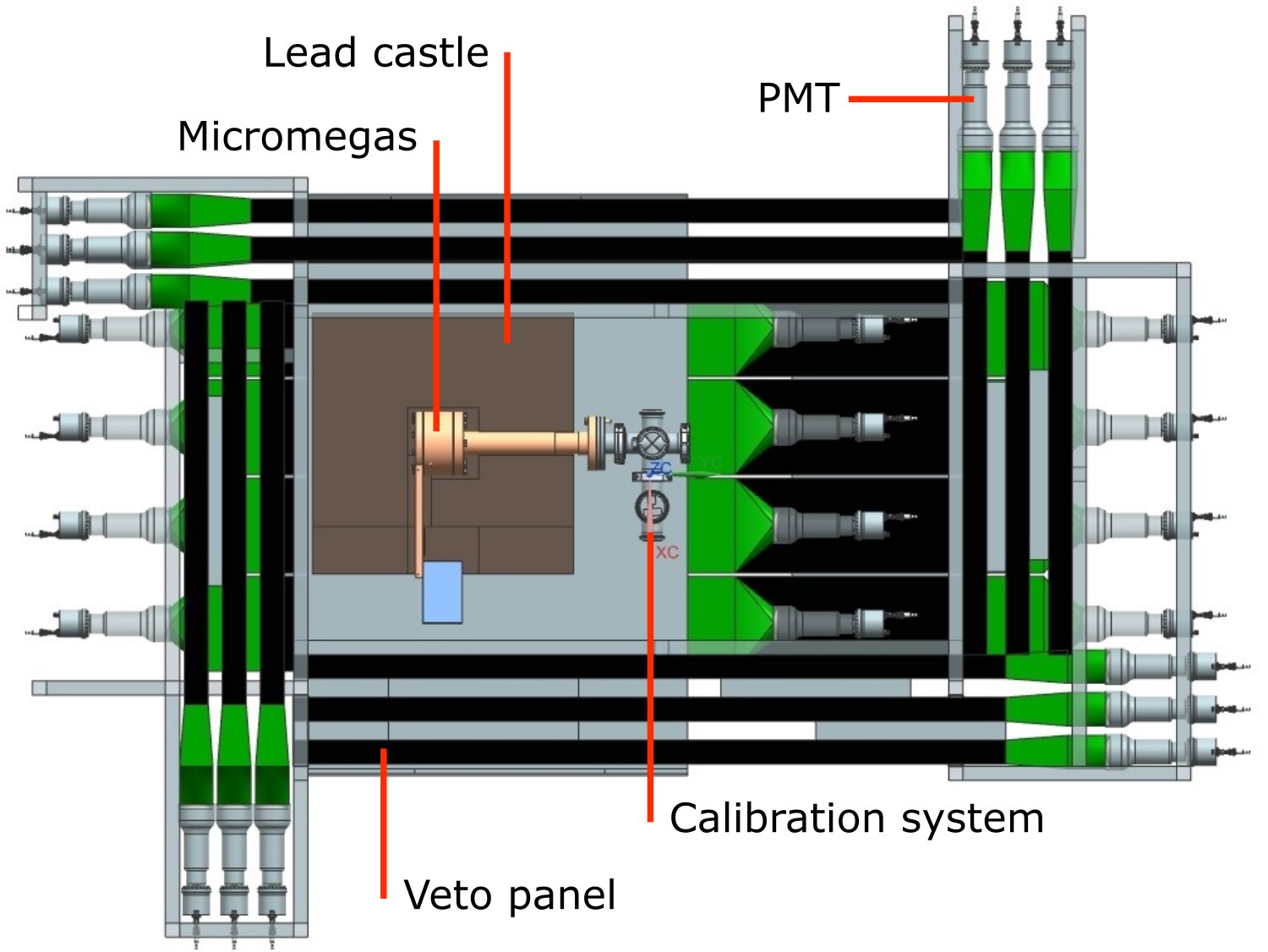}
		\caption{A cutaway of the setup with the triple-layer veto system, seen from the top. The cadmium sheets surrounding the veto panels are not pictured.}
		\label{fig:veto_sketch}
	\end{figure}

	In total 57 veto panels were installed. A drawing of the system is shown in figure \ref{fig:veto_sketch}. The veto panels are made of NE-110. They are second-hand from the 90s and were previously used in a time-of-flight spectrometer \citep{Scintillators}. For our setup, the 3\,m long panels were cut to the appropriate length of 150\,cm or 65\,cm. Each panel has a width of 20\,cm and thickness of 5\,cm. On one end a PMT is connected via a light guide. In between the panels, cadmium sheets of 1\,mm thickness were placed.
	The veto panels were calibrated using the spectrum of atmospheric muons. The energy loss of muons in matter follows a Landau distribution, where the most probable value -- the peak -- is located at 
	2\,MeV/g\,cm\textsuperscript{2}. Thus, if the veto panel lies flat, with its 5\,cm thickness and a density of $\sim 1$\,g/cm\textsuperscript{3} the peak corresponds to 10\,MeV.
	It has to be kept in mind that the light attenuation over the length of the scintillators is significant. Measurements of a few panels showed that at the far end of a 150\,cm panel the light is attenuated by a factor of two. Thus the calibrated energy of a veto signal can only serve as a lower limit of the real energy.
	
	\subsection{Electronic readout}
	Both the Micromegas and the vetoes are read out by 4 AGET ASICs each~\citep{AGET}. Each AGET has 64 channels. The shaped detector signals are digitized and stored over an adjustable time window. The analog signal is sampled continuously over 512 bins by a  circular buffer. This allows to freely select the delay between the start of the time window and the trigger. In order to use different readout settings for the Micromegas and the vetoes we use separate FPGAs for each group of AGETs. A central trigger and clock module synchronizes the two groups of AGETs. For the Micromegas, the time window should be large enough to collect entire tracks over the 3\,cm drift distance. With typical voltage settings, the drift velocity of electrons in the Xe-Ne mixture is about 5\,mm/$\upmu$s, thus an event can take up to 6\,$\upmu$s. During data taking, the time window was set to different values between 10 and 30\,$\upmu$s.
	The veto readout system is triggered by the Micromegas readout electronics. Whenever the Micromegas registers an event, all veto channels are read out in a 100\,$\upmu$s time window. In relation to this window, the Micromegas trigger is located at 30\,$\upmu$s.
	
\section{Simulations}
\label{sec:sims}

The development of the experiment has been complemented by extensive radiation transport simulations using Geant4~\citep{Geant4_2016}. REST-for-Physics has been used as the software framework to interface with Geant4, and to process the resulting data~\citep{REST}. 
REST-for-Physics produces a full simulation of the detector: the energy deposits calculated by Geant4 are processed to mimic raw signals similar to the ones measured by the physical detector, allowing for an accurate comparison. The conversion takes into account different physical processes taking place in the detector such as the electron drift and diffusion.
Special attention has been given to the simulation of the veto system. The framework has been enhanced to allow the full simulation of such a system producing raw signals for the veto system after also simulating processes such as light quenching or light attenuation as measured in the physical setup.

Different backgrounds have been simulated, in particular the cosmic-ray induced neutron flux at sea level~\citep{COSMIC_NEUTRON_FLUX}. Cosmogenic neutrons are of special interest because they can often produce events with a signature similar to axion-induced events. At the same time it is very difficult to shield the detector from this background contribution at surface level so a strategy to tag these events was developed.
Simulation studies for cosmic-ray induced neutrons result in a raw background level of $8.8 \times 10^{-6} \, \textnormal{counts keV}^{-1} \, \textnormal{cm}^{-2} \, \textnormal{s}^{-1}$ on the readout surface for a ROI of 1--10\,keV. This background level is further reduced after applying micromegas signal cuts, described in~\ref{sec:XrayCuts}, to a value of $9.0 \times 10^{-7} \, \textnormal{counts keV}^{-1} \, \textnormal{cm}^{-2} \, \textnormal{s}^{-1}$. The background level is strongly dependent on the primary neutron flux assumed in its calculation, which currently has a large uncertainty. In any case, it supports the hypothesis that the cosmogenic neutrons are the dominating contribution to the remaining background level achieved e.g. with the Micromegas at CAST, motivating the development of the active shielding system described above, able to tag neutron-induced events. 

As already described in section \ref{sec:setup}, the muon veto system was used as the base for the cosmic neutron veto. To enhance the capability of detecting cosmic neutrons, the number of plastic scintillators was increased to achieve a three-layer 4$\pi$-coverage and cadmium sheets were introduced as neutron capturing material around the scintillators (see figure \ref{fig:veto_sketch}).

\begin{figure}
    \centering
    \includegraphics[width=\textwidth]{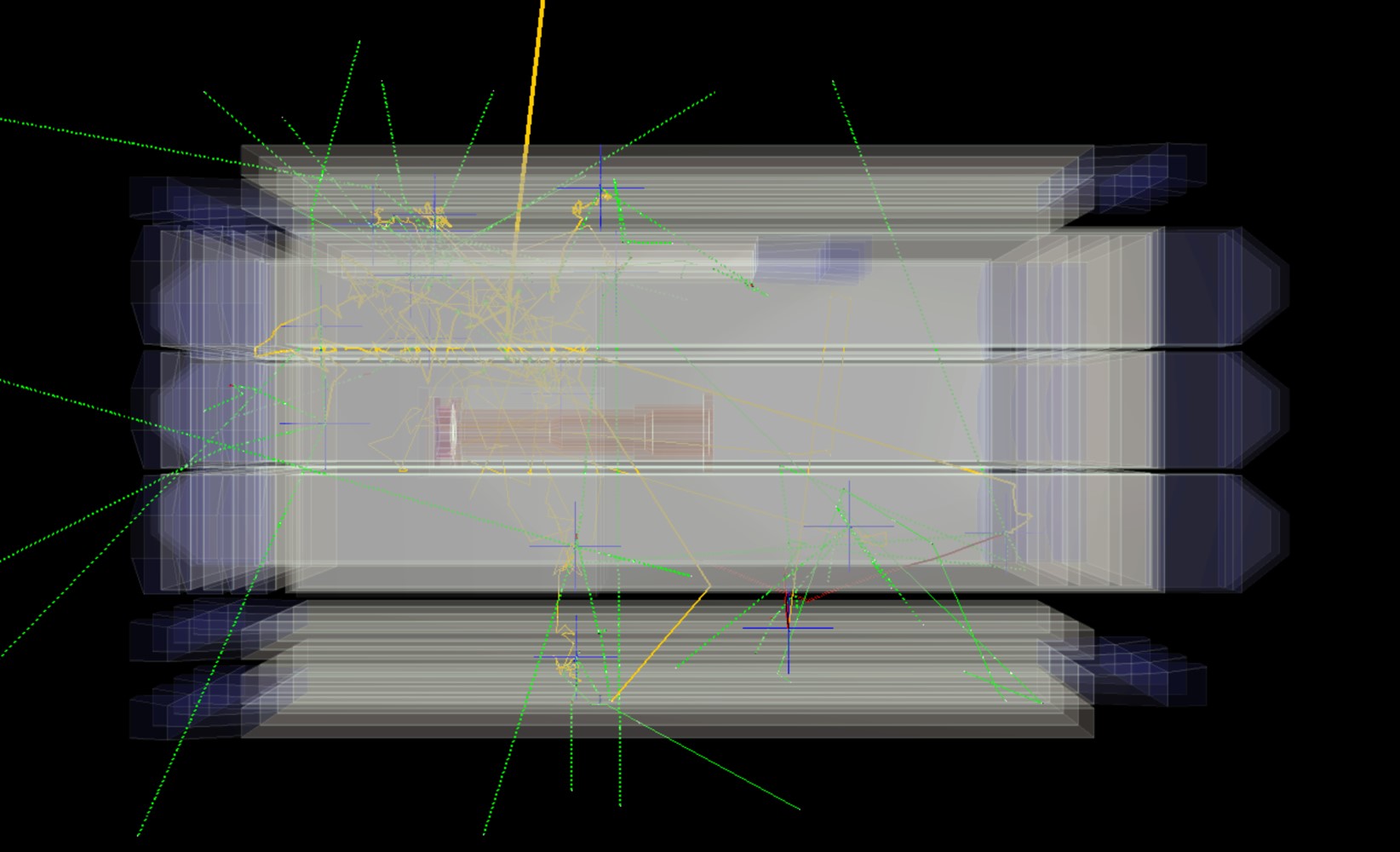}
    \caption{Simulation of a cosmic neutron event. The primary neutron (thick yellow line) enters the setup from above interacting in the lead castle producing multiple secondary lower-energy neutrons (yellow). Some of these neutrons interact with the veto system and after moderation can be captured (blue cross), resulting in the production of gamma rays (green) which can be detected by the veto.}
    \label{fig:simulation_neutron_event}
\end{figure}

An example of a simulated cosmic neutron event is displayed in figure \ref{fig:simulation_neutron_event}. All cosmic neutron events follow a similar pattern: a high energy neutron enters the setup from above usually interacting with the lead shielding. These interactions produce a number of secondary neutrons roughly proportional to the primary energy. These secondary neutrons can also multiply if their energy is high enough. The lower energy secondary neutrons gradually lose energy via elastic processes in the shielding and especially in the plastic scintillators, which  consist of elements with low atomic number.
These direct interactions can produce signals in the Micromegas, but also in the veto panels. However, the recorded light yield is quenched compared to photon or electron interactions.
After losing most of their kinetic energy, the neutrons are captured by the cadmium sheets. The result of this capture is a burst of gamma rays with energies in the order of a few MeV. Since this capture takes place near the scintillators, the gammas produced could be used to improve the efficiency of the veto system. The Micromegas itself is protected against these gammas by the lead shielding.

\begin{figure}
    \centering
    \includegraphics[width=1.0\textwidth]{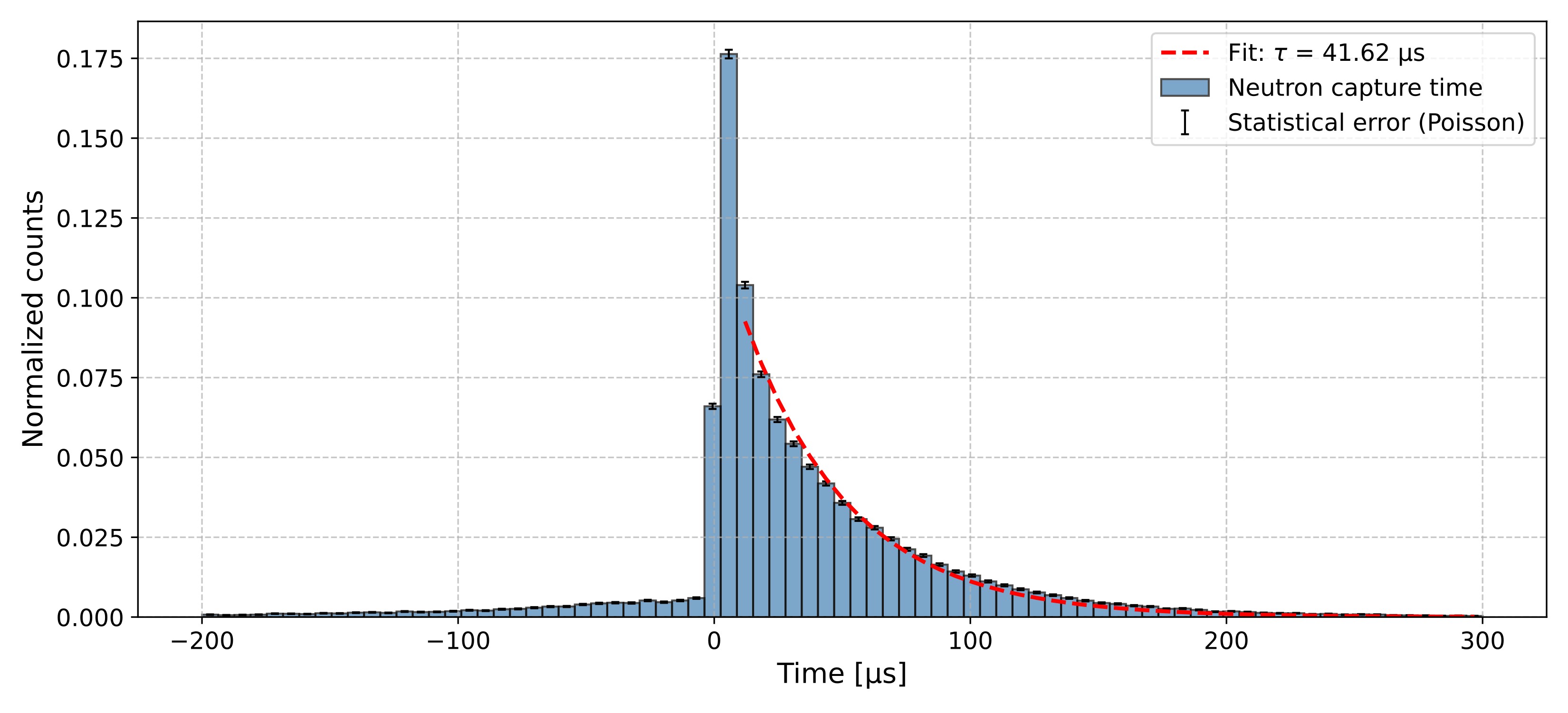}
    \caption{Time of neutron capture in the simulations for cosmic neutron background events. Here $t=0$ corresponds to the trigger of the Micromegas detector. A fit to a negative exponential function is displayed.}
    \label{fig:neutron_capture_time}
\end{figure}


\begin{figure}
    \centering
    \includegraphics[width=1.0\textwidth]{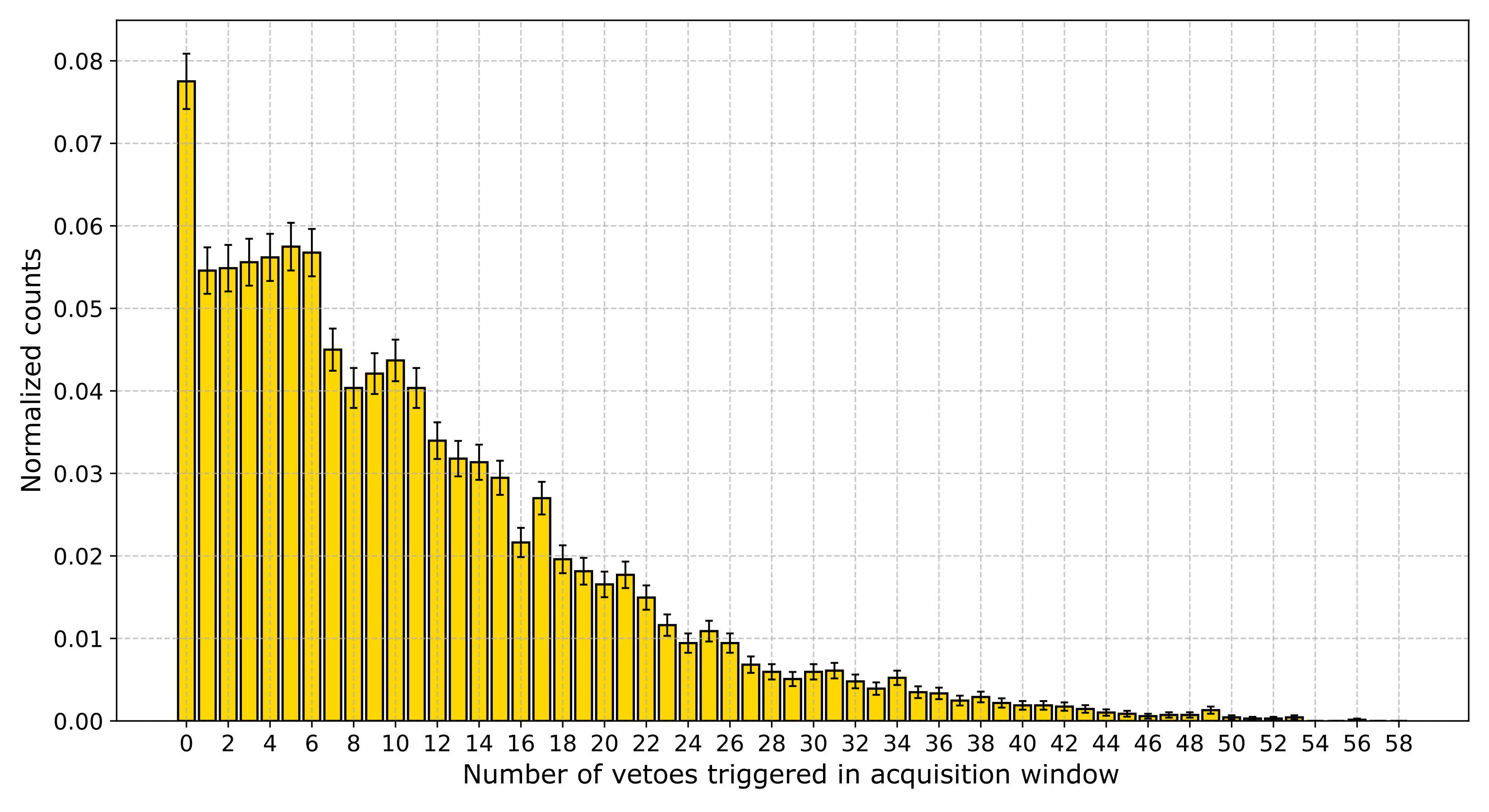}
    \caption{Simulation of the multiplicity -- the number of triggered vetoes -- for a neutron event. The veto threshold is set to 1\,MeV. The counts are shown with their statistical errors.}
    \label{fig:neutron_simulation_multiplicity}
\end{figure}

Due to the secondary neutrons needing to become thermalized before being captured, there is a delay between the background event in the Micromegas and the multiple neutron captures, as shown in figure~\ref{fig:neutron_capture_time}. One can see that in order to detect most captures a time window in the order of 100\,$\upmu$s is required.
Figure \ref{fig:neutron_simulation_multiplicity} shows the multiplicity of veto signals of a neutron event. The entire chain from the production of secondary neutrons to the interaction of single gammas in multiple veto panels was taken into account. The threshold of the vetoes was set to 1\,MeV. The bin at 0 corresponds to events, were no neutron was captured. One can see that in most events many vetoes detect a signal, with the multiplicity distribution peaking in the range of 1 to 6, while it steeply descends for higher values.

	\section{Data processing \& analysis}
	
	\subsection{Data processing with REST-for-Physics}
	The data acquired with the Micromegas detector is processed and analyzed using the REST-for-Physics software framework~\citep{REST}. This software processes the signals from the strips and extracts spectra and 3-dimensional topological information of the charge deposition.
	The data is processed in three steps using the appropriate REST-for-Physics libraries: during the raw signal analysis noise events are identified and removed, and the signal pulses are isolated; in the detector hits analysis the signals from electronic channels are translated to energy deposits (``hits") and associated with physical coordinates in the readout geometry; finally, in the track analysis the energy deposits are connected to 3-dimensional tracks to add another level of topological information to each event. The processing generates many observables -- per-event information extracted from the data -- that are used to define selection criteria for background discrimination.
		
	\subsection{Background discrimination using the Micromegas data}
    \label{sec:XrayCuts}
	The signature of an axion-like signal is determined by calibrating the detector with the 5.9\,keV X-rays of the \textsuperscript{55}Fe-source. In general, a low energy X-ray is expected to leave a small, symmetrical charge deposit in the detector. Furthermore, in the context of BabyIAXO, the signals of interest are located in a small spot in the center of the detector readout plane -- the focal spot of the telescope. Also the calibration events in the IAXO-D0 setup are centered in a small region of the detector. This collimation is caused by the source distance and the size of the entrance window.
 
\begin{figure}
		\centering
		\includegraphics[width=\textwidth]{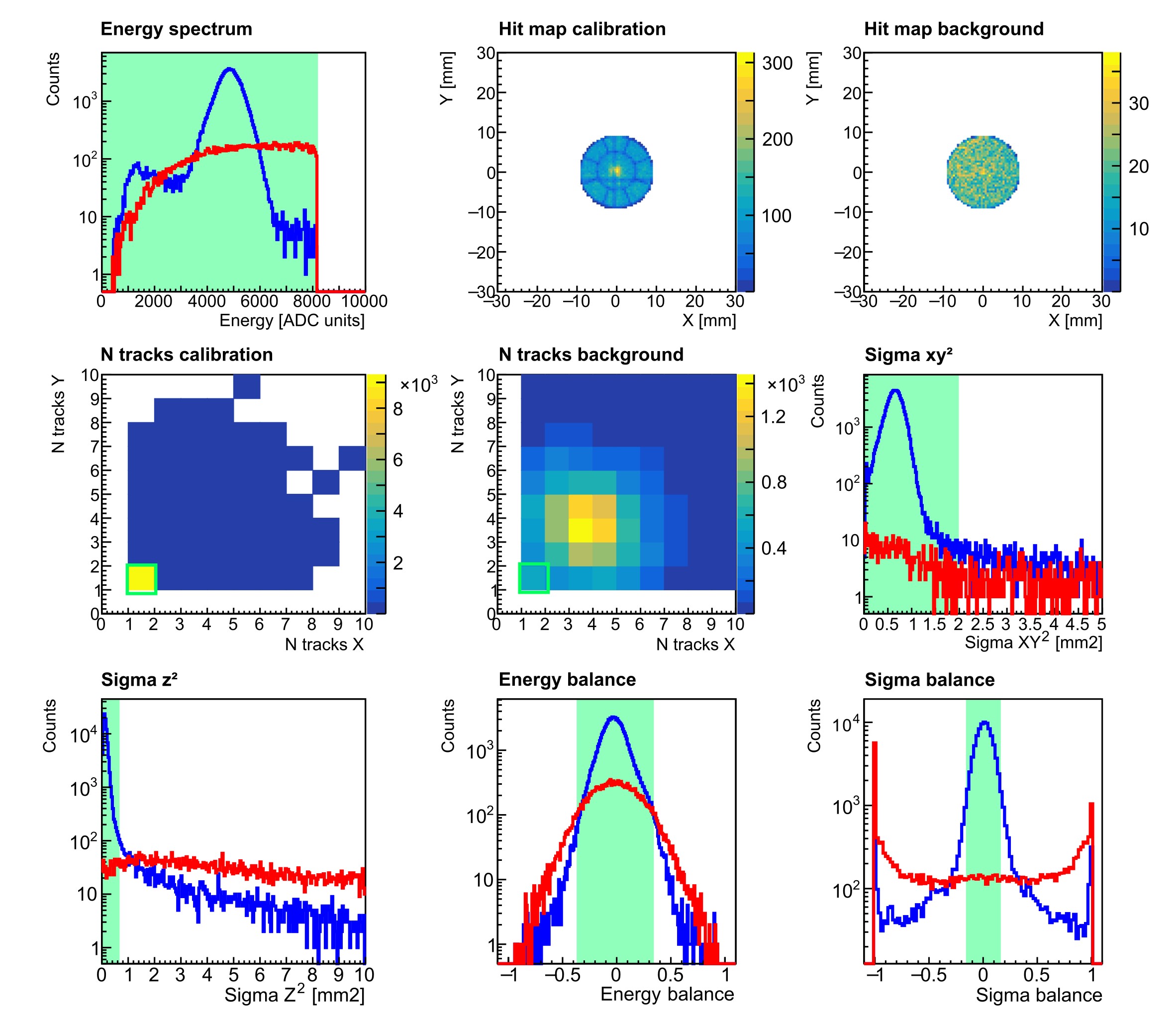}
		\caption{Comparison of data obtained with the \textsuperscript{55}Fe calibration-source (blue histograms) and background runs (red histograms). The green markings indicate the selection criteria, all events outside the shaded areas are removed from the data.
        The energy and fiducial cut were already applied. For the number of tracks observable, the best discrimination is achieved by selecting only single-track, i.e. point-like events. Furthermore, as shown in the ``sigma'' observables, events with small spread in all spatial coordinates are selected. For the balance observables, we select events in a small interval around zero, corresponding to symmetric energy deposits.}
		\label{fig:cuts}
	\end{figure}

	The following observables were used for background discrimination:
	\begin{itemize}
		\item Energy: the cuts are optimized using the 5.9\,keV peak of the \textsuperscript{55}Fe-source. Events within about one sigma around the peak maximum are used, typically corresponding to an energy range of $5.4 - 6.4$\,keV
		\item Position in the readout (X--Y) plane: only events where the mean position of the deposited energy of an event falls within a circle of 9\,mm radius in the center of the readout plane are considered. The size of this circle is determined by the spot of the collimated X-rays from the calibration source.
		\item Number of tracks: the photoabsorption of X-rays is a point like event that is identified as a single track in contrast to background events like charged particle interactions that usually leave multiple tracks.
		\item $\sigma_{XY}^{2}$: sum of variance of the hits in X and Y coordinates, i.e. the size of the event in X and Y.
		\item Sigma balance in X--Y plane: the sigma balance $(\sigma_{X} - \sigma_{Y})/(\sigma_{X} + \sigma_{Y})$ measures the topological symmetry of event.
		\item Energy balance: the energy balance $(E_{X} - E_{Y})/(E_{X} + E_{Y})$ is rather similar to the sigma balance, but uses the reconstructed energy.
		\item $\sigma_{Z}^2$: the variance of the event in the Z-coordinate.
	\end{itemize}
	To discriminate background, the data is filtered using the above-mentioned observables. Each cut is either a function of one variable (e.g. the radius of the focal spot) or of two variables (upper and lower limit). The variables depend on the conditions of the measurement and change for example with the detector voltage settings, which affect the diffusion and spread of events.
	The variables have to be optimized such that the filter removes as much as possible background events while preserving calibration data. This optimization is performed via an automatic procedure using the MINUIT2 engine by maximizing the ratio of calibration-to-background events for each cut sequentially.
    In figure \ref{fig:cuts} a dataset is plotted and the applied criteria are shown.

	\subsection{Background discrimination with the veto system}
\begin{figure}
		\centering
		\includegraphics[width=0.6\textwidth]{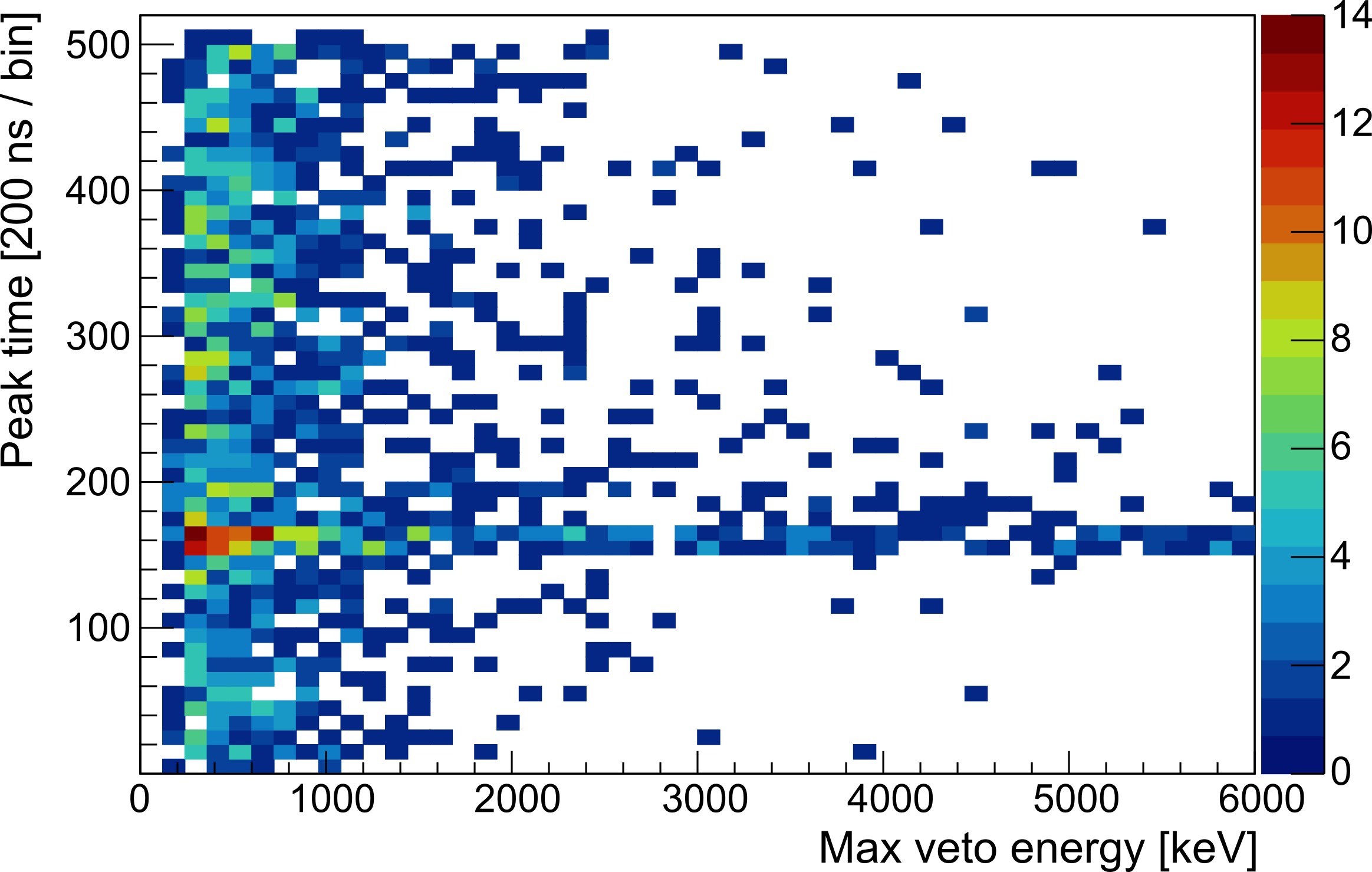}
		\caption{A histogram of the veto signals from background data with their energy and peak time. The excess at peak time bin $\sim 150$ are muons in coincidence with a Micromegas signal. Note that every event can contain up to 57 signals in the vetoes (one for each panel), which are displayed here.}
		\label{fig:veto_data}
	\end{figure}

\begin{figure}
		\centering
		\includegraphics[width=0.49\textwidth]{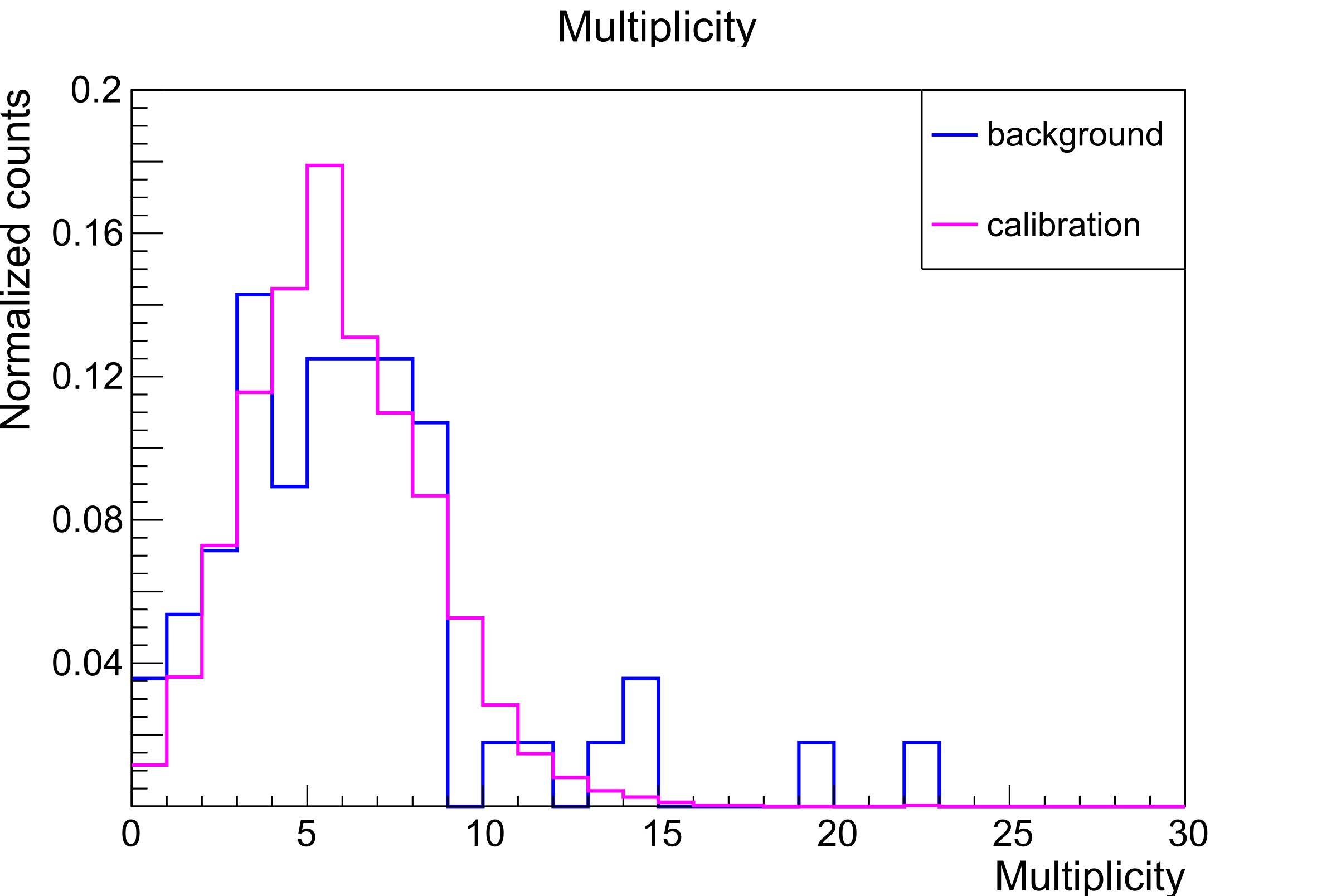}
		\includegraphics[width=0.49\textwidth]{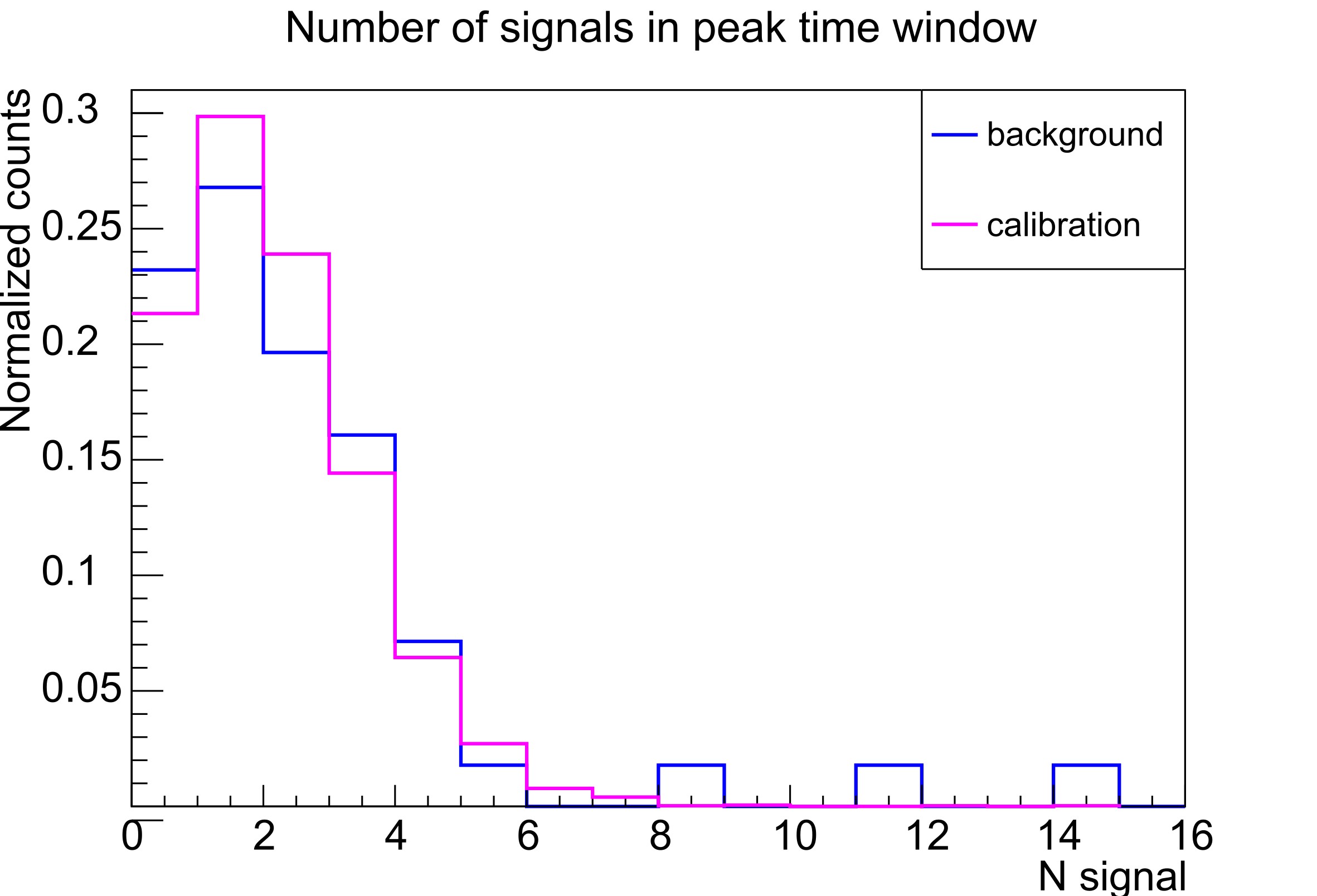}
		\includegraphics[width=0.49\textwidth]{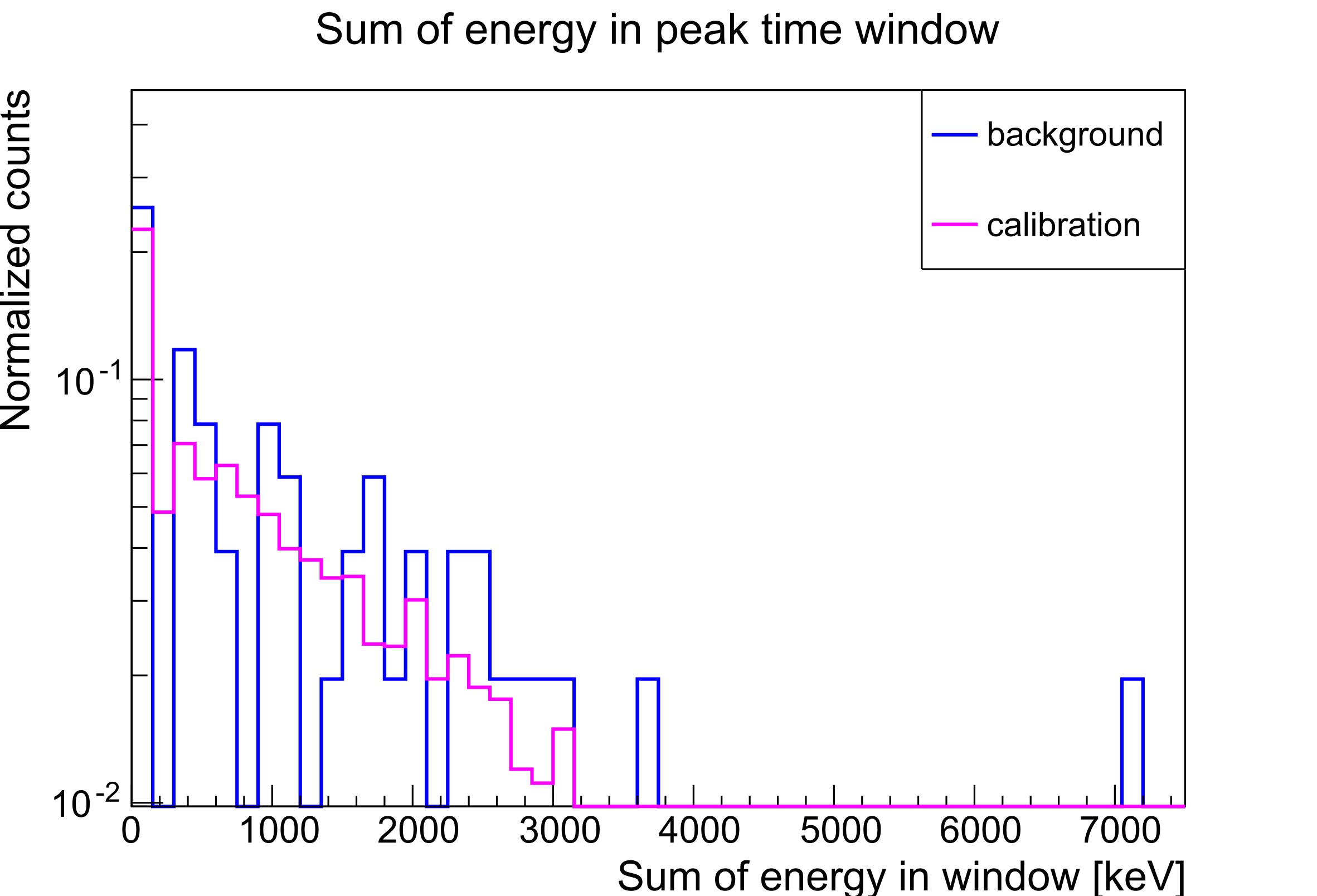}
		\includegraphics[width=0.49\textwidth]{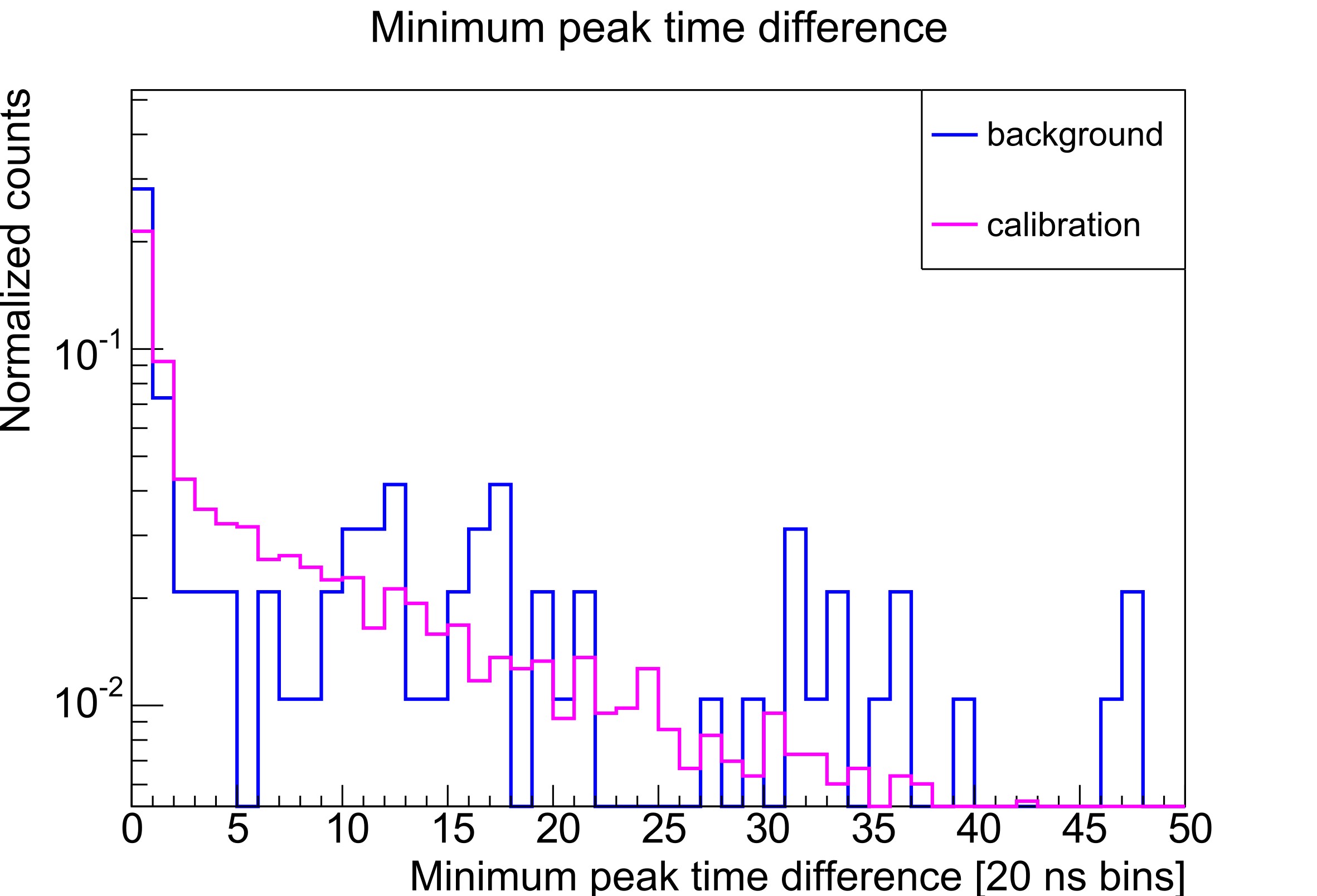}
		\caption{The observables generated from the veto data to discriminate background. The multiplicity (overall and in a reduced time window) and sum of energies have an excess of signals at higher values for the background data. The plot on the lower right shows the minimum time difference between two signals of an event: 0 means that at least two signals are in coincidence, which is more common in the background data.}
		\label{fig:neutron_cuts}
	\end{figure}

	A muon cut is applied using data from the veto system. The relevant time window can be easily identified by observing an excess of signals in temporal correlation with the Micromegas trigger. This can be seen in figure \ref{fig:veto_data}. The muon cut removes events which contain a veto signal within a time window of 16\,$\upmu$s around the Micromegas trigger and with an energy of more than 5\,MeV.
	After the analysis of the data of the veto system, additional non-muon background can be discriminated. As shown in section \ref{sec:sims}, the signature of background induced by cosmic neutrons is expected to be events with a high multiplicity, with the individual signals spread over several tens of microseconds. Additionally, direct neutron interactions in the vetoes or multiple gamma interactions after the capture can result in coincident signals in multiple vetoes. 
    It was found that the following observables facilitate to identify background events:
	\begin{itemize}
		\item Overall multiplicity: number of hit vetoes per event.
		\item Multiplicity within a time window: the number of veto signals within a fixed time window.
		\item Sum of energy of veto signals within a time window.
		\item Minimum peak time difference of the signals in an event: if this value is small or 0, it means that two or more signals came in coincidence.
	\end{itemize}
	The parameters for these cuts were optimized using the optimization procedure described above, with independent time windows for each. Histograms of these observables are shown in figure \ref{fig:neutron_cuts}. In general it can be observed that the multiplicity and sum of energies observables show an excess of signals at higher values in the background data.
	
	\section{Results}
	
	The effective duration of the analyzed background data is 52.1\,days. In this time 1,305,996 raw events were collected in the entire detector and full energy range. By restricting the energy to the ROI of 2--7\,keV and discarding events outside the central spot with 9\,mm radius, the number of events is reduced by more than 90\%.
	Applying the other selection criteria described in section \ref{sec:XrayCuts} (also see figure \ref{fig:cuts}) using only the Micromegas data, 257 background events remain (0.02\%). In the calibration data most events pass the cuts at an efficiency of 81.9\% in the ROI.

	The muon discrimination with the veto system further reduces the number of background events by a factor of 4.6, down to 56 events. Thus the background level in the focal spot and ROI is $9.78 \times 10^{-7} \, \textnormal{counts keV}^{-1} \, \textnormal{cm}^{-2} \, \textnormal{s}^{-1}$. This value is comparable with previous results reported in \citep{CAST}.

\begin{figure}
        \centering
        \includegraphics[width=0.6\textwidth]{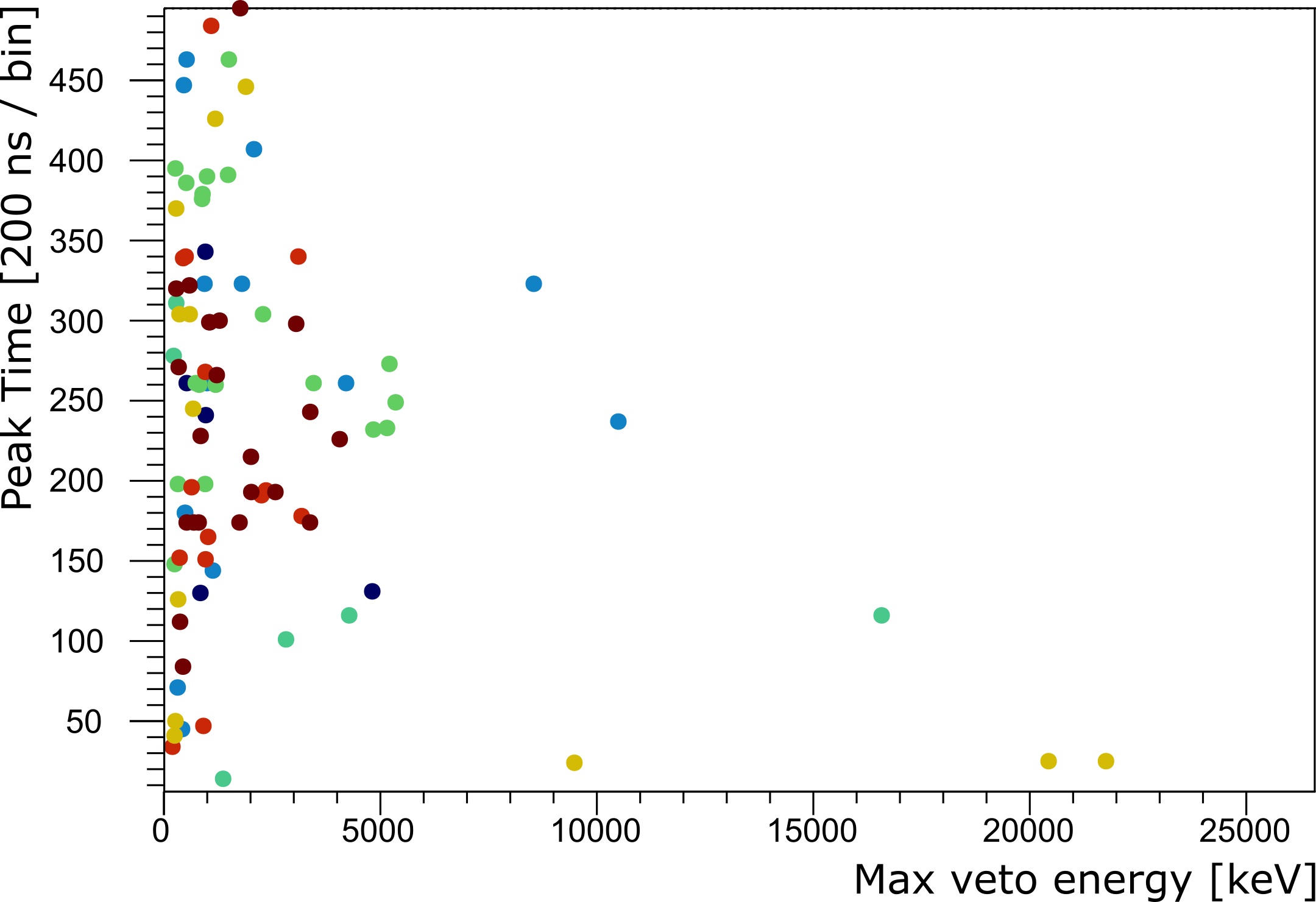}
        \caption{This plot shows the energy and peak time of the veto signals that were identified as background, using the advanced discrimination after the muon cut. The color represents the event, i.e. all points with the same color belong to the same event.}
        \label{fig:cut_events}
    \end{figure}

	This result is improved further with the advanced filter using the veto system. The best results were obtained with following conditions: overall multiplicity $< 13$; sum of energy $< 7034$\,keV in the time window [23,53]\,$\upmu$s; multiplicity $< 12$ in the time window [35,68]\,$\upmu$s. 
    In this data the observable of coincident events in the vetoes was redundant with the others, but in general also allows background discrimination.
    In summary, the signature of these events are a high multiplicity, with many signals clustered in a time window of $\sim 30\,\upmu$s that is offset to later times with respect to the Micromegas trigger. Additionally, some signals are detected in coincidence. 
    The background events identified by this cut are shown in figure \ref{fig:cut_events}. In this plot, the trigger is located at peak time bin 150. One can see that most signals are spread after the Micromegas trigger and have low energies close to the threshold.
 
\begin{figure}
		\centering
		\includegraphics[width=0.6\textwidth]{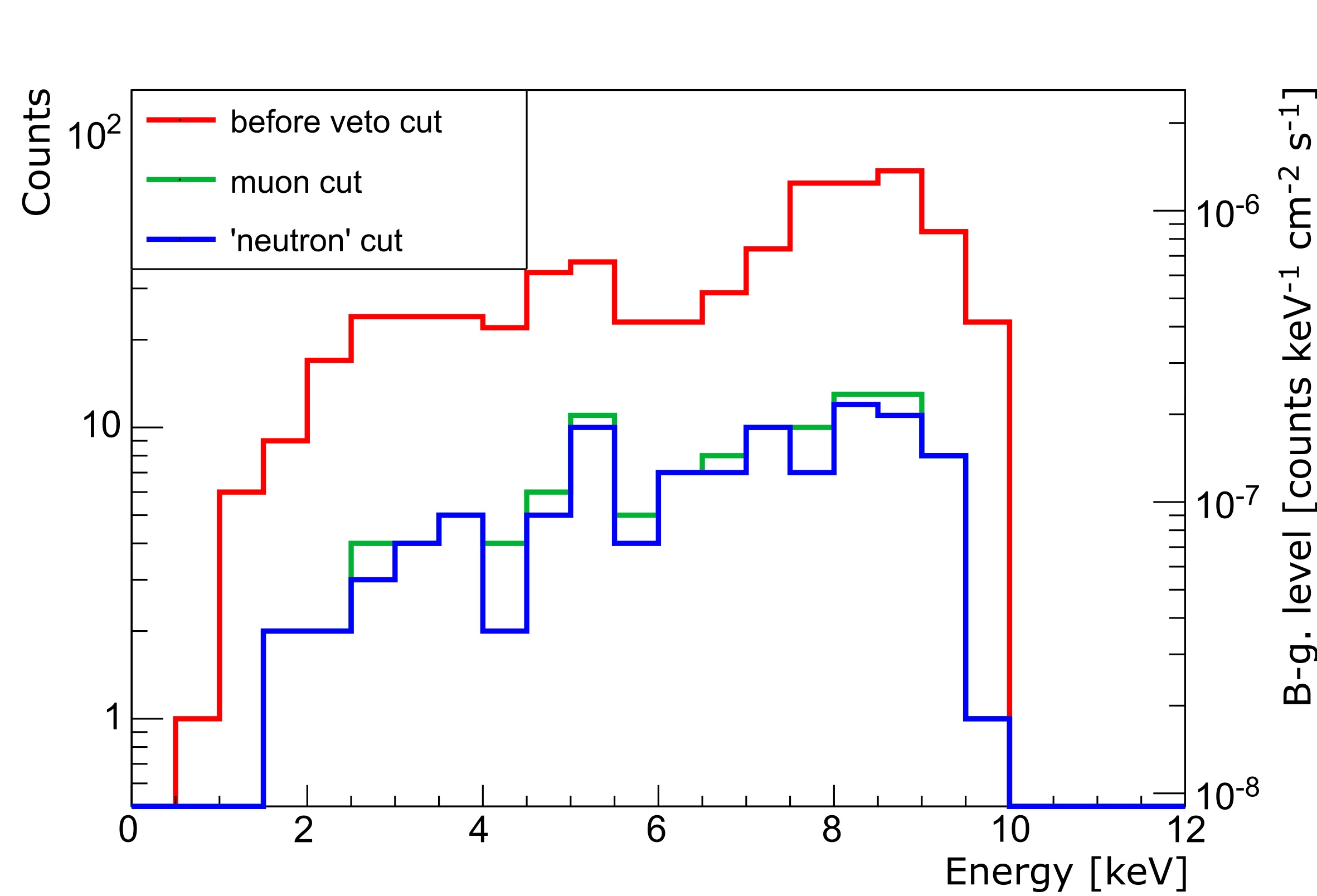}
		\caption{The spectrum of X-ray like events in the Micromegas below 10\,keV, before the muon cut (red), after removing muons (green) and after applying the advanced cuts using the veto (blue). On the left axis, the number of counts is shown, while on the right the normalized background level is indicated. The integrated background level is calculated in the range of interest of $2-7$\,keV to avoid the copper X-ray fluorescence peak at 8\,keV.}
		\label{fig::MMspectrum}
	\end{figure}

	These advanced cuts reduce the number of background events in the central spot and in the ROI by 13\% , down to 49 events, while 97\% of calibration events are preserved. In summary, and including the statistical error, a background level of $$(8.56 \pm 1.22) \times 10^{-7} \, \textnormal{counts keV}^{-1} \, \textnormal{cm}^{-2} \, \textnormal{s}^{-1}$$ is reached, with an efficiency of 79.4\% in the calibration data within the ROI of 2--7\,keV.
    The spectrum of the Micromegas signals before and after the veto cuts is shown in figure \ref{fig::MMspectrum}.


	\section{Conclusions and future developments}
	The background level reported here is the lowest ever achieved at above ground level with this type of detector. Comparing the measurements with the simulations presented in the paper, one can see that the measured background rate is at the same level as the simulated neutron-induced background. This indicates that still a large part of this background contribution is not discriminated by the multi-layer veto system. However, it has been shown that looking specifically for events with a neutron-like signature in the vetoes -- a high signal multiplicity spread over several tens of microseconds -- improves the background level by 13\,\% compared to the one after the muon cut. The veto signals of these additionally rejected events have low energies. Thus we expect that a veto system with a lower energy threshold and a higher segmentation would improve the identification of neutron-induced background events. One has to keep in mind that the current veto was constructed with readily available materials, re-using and adapting several decade-old scintillator panels. An upgrade would be realized with new equipment specifically designed for BabyIAXO.

    In general, the goal of the detector development is to reduce the background level even more to $1 \times 10^{-7} \, \textnormal{counts keV}^{-1} \, \textnormal{cm}^{-2} \, \textnormal{s}^{-1}$. To achieve this, two new Micromegas prototype setups have been prepared, one at surface level at the University of Zaragoza, the other in the Canfranc underground laboratory (LSC). This setup will help to understand the contributions of cosmogenic and terrestrial background sources better. In addition, the radiopurity of the components of the new detectors has been improved, and radiopure electronics are in development. Also the calibration capabilities have been enhanced by using custom X-ray generators. The setup at LSC is already in operation, while the other is currently under commissioning. Additionally, the collaboration is working on an improved veto-system.

    \section*{Acknowledgments}
	
 This work has been performed as part of the IAXO collaboration. We would like to thank our collaborators, in particular the ones from the IAXO Detector Working Group. We also thank in particular the INR and U. Mainz colleagues for enabling the in-kind contribution consisting of the plastic scintillators used in this work. 
 The authors would like to acknowledge the use of Servicio General de Apoyo a la Investigación-SAI, Universidad de Zaragoza.
 
We acknowledge support from the European Union’s Horizon 2020 research and innovation programme under the European Research Council (ERC) grant agreement ERC-2017-AdG788781 (IAXO+)
and under the Marie Skłodowska-Curie grant agreement No 101026819 (LOBRES); grant PID2019-108122GB-C31 funded by MCIN/AEI/10.13039/501100011033 and funds from “European Union NextGenerationEU/PRTR” (Planes complementarios, Programa de Astrof\'isica y F\'isica de Altas Energías). We also acknowledge support from the Agence Nationale de la Recherche (France) ANR-19-CE31-133 0024.

	\bibliographystyle{Frontiers-Harvard}
	\bibliography{bibliography}
	
\end{document}